\documentstyle[12pt]{article}
\def\c60{~C$_{60}$}
\def\be{\begin{equation}}
\def\ee{\end{equation}}
\def\bea{\begin{eqnarray}}
\def\eea{\end{eqnarray}}
\def\bq{{\vec q}}
\def\br{{\bf r}}
\def\Oh{{\hat \Omega}}
\def\half{\frac{1}{2}}
\begin{document}
\baselineskip 18pt
\title{Electron--Vibron Interactions
and Berry Phases in Charged
Buckminsterfullerene: Part I}
 \author{Assa Auerbach\thanks{Email: assa@phassa.technion.ac.il}\\
{\it Physics Department, Technion, Haifa, Israel},\\
Nicola Manini$^a$\thanks{Email: manini@tsmi19.sissa.it} and
Erio Tosatti$^{a,b}$\thanks{Email: tosatti@tsmi19.sissa.it},\\
{\it a) International School for Advanced Studies (SISSA),}\\
{\it via Beirut 2-4, I-34013 Trieste, Italy.}\\
{\it b) International Centre for Theoretical Physics (ICTP),}\\
{\it P.O. BOX 586, I-34014 Trieste, Italy.}}
\maketitle

\begin{abstract}
A simple model for electron-vibron interactions on charged buckminsterfullerene
\c60$^{n-}$, $n=1,\ldots 5$, is solved both at weak and strong couplings.  We
consider a single $H_g$ vibrational multiplet interacting with $t_{1u}$
electrons. At
strong coupling  the  semiclassical  dynamical Jahn-Teller theory is valid.
The Jahn-Teller distortions are unimodal for $n$=1,2,4,5 electrons, and
bimodal for 3 electrons. The distortions are  quantized as rigid body
pseudo--rotators which are subject to geometrical Berry phases. These impose
ground state degeneracies and dramatically change zero point energies.  Exact
diagonalization shows that the semiclassical level degeneracies and ordering
survive well into the weak coupling regime.  At weak coupling, we discover an
enhancement  factor of $5/2$ for the pair binding energies over their classical
values. This has potentially important implications for superconductivity in
fullerides, and demonstrates the shortcoming of Migdal--Eliashberg theory for
molecular crystals.
\end{abstract}

\section{Introduction}
\label{intro}

The soccer-ball shaped molecule C$_{60}$ (buckminsterfullerene)  and its
various crystalline compounds  have ignited enormous interest in the chemistry
and physics community in past two years \cite{hebard}. C$_{60}$ is a truncated
icosahedron. From a physicist's standpoint, the charged molecule is
fundamentally interesting, because the high molecular symmetry gives rise to
degeneracies in both electronic and vibrational systems. Thus, the  molecule is
very sensitive to perturbations. In particular, electron--phonon and
electron--electron interactions are expected to produce highly correlated
ground states and excitations.

Superconductivity has been discovered in alkali doped
buck\-min\-ster\-ful\-le\-ren\-es A$_3$C$_{60}$ (A=K,Cs,Rb), with  relatively
high transition temperatures ($T_c\approx20^\circ$--$30^\circ K$). There are
experimental indications that the pairing mechanism originates in the
electronic properties of a single molecule. The pair binding energy is a
balance of electron-vibron interactions \cite{vzr,sch,t1} and electron-electron
interactions \cite{ck}. The relative contributions and signs of the two
interactions is under some controversy.

The electron--vibron school has identified certain five-fold degenerate $H_g$
(d-wave like) vibrational modes which couple strongly to the $t_{1u}$ Lowest
Unoccupied Molecular Orbital (LUMO) \cite{vzr,sch,t1,johnson}. Varma, Zaanen
and Raghava\-cha\-ri \cite{vzr} as well as Schluter {\em et al} and, more
recently, Antropov {\em et al} proposed that these modes undergo a Jahn-Teller
(JT) distortion and calculated the induced pair binding energies at several
fillings.  They used the {\em classical approximation}, and restricted their
calculation to unimodal distortions (defined later). The general conclusion of
this approach is that, while the calculated $\lambda$ is sizeable, one still
requires a large reduction of the Coulomb pseudo--potential $\mu^*$ in order to
explain the highest transition temperatures. On the other hand, Gunnarsson {\em
et al} independently estimate a large $\mu^* \approx 0.4$, i.e. there is no
mechanism providing such a reduction.

However, estimates of the electron--vibron coupling constant $g$ do not justify
the classical JT approximation. C$_{60}$ is estimated by frozen phonon
calculations to be in the weak coupling regime $g\le 1$ where quantum
corrections are important.

In this paper (Part I) we study the isolated C$_{60}^{n-}$ charged molecule. In
particular, we shall reconsider the same JT model, but diagonalize the
quantum Hamiltonian for  the full range of the coupling constant. We shall find
that quantum corrections to the classical JT theory introduce novel qualitative
features, and are quantitatively important for the pair binding energies.

The quantum fluctuations involve interference effects due to geometrical Berry
phases. Berry phases appear  in a wide host of physical phenomena
\cite{mead,w}. Here we find it in the context of a ``Molecular Aharonov-Bohm
(MAB) effect'', originally discovered by Longuet-Higgins \cite{lh}. The MAB
effect has important consequences on the vibron spectrum. For example, it
produces  half--odd integer quantum numbers in the spectrum of triangular
molecules \cite{lh,mead}, an effect recently confirmed spectroscopically in
Na$_3$ \cite{delac}. This kind of Berry phase is important also in scattering
of hydrogen molecules \cite{levi}. Recently, it has been suggested that a
geometrical Berry phase may be relevant in fullerene ions
\cite{tosatti,gonzalez,ihm}.  Here we show that Berry phases produce {\em
selection rules} for the  pseudo--rotational quantum numbers and {\em
kinematical} restrictions which effect the pairing interaction between
electrons.  Although the semiclassical and Berry phase description is
appropriate in strong coupling, the level ordering and degeneracies are found
to survive for arbitrary coupling, particularly in the weak coupling regime,
which is closer to actual C$_{60}$. For this reason we devote a large portion
of this paper to the semiclassical theory, which helps to build physical
intuition for further extensions of the model.

This paper is organized as follows: In Section \ref{model} the basic model is
introduced. Section \ref{class} calculates the JT distortions in the
classical limit. Section \ref{semiclas} derives the semiclassical quantization
about the JT manifold. The geometrical Berry phases are calculated,
and their effect on the semiclassical spectrum is obtained up to order
$g^{-2}$.   Section \ref{exact} describes the exact diagonalization results,
and compares them to the semiclassical theory, and to weak coupling
perturbation theory. The pair binding energies  are determined in Section
\ref{pair-bind}.  In Section \ref{discuss} we summarize the paper and discuss
our main result: that the effective pair binding energies are larger by a
factor of $3$ than the pair  interaction energy in Migdal--Eliashberg's theory.
In a following paper \cite{II} we shall extend the model to  all $A_g$ and
$H_g$ modes with realistic physical parameters. This will allow us to explore
the experimental consequences of the electron--vibron interactions.

\section{The  Electron--Vibron Model}
\label{model}
The single electron LUMO states of \c60 are in a triplet of $t_{1u}$
representation. We consider the $H_g$  (five
dimensional) vibrational multiplet which couples to these
electrons. $t_{1u}$ and $H_g$ are the icosahedral group counterparts of the
spherical harmonics  $\{Y_{1m}\}_{m=-1}^1$, and  $\{Y_{2M}\}_{M=-2}^2 $
respectively. By replacing  the truncated icosahedron (soccer ball) symmetry
group by the spherical group, we ignore lattice corrugation effects. These are
expected to be small since they do not lift
the  degeneracies of the $L=1,2$ representations.

The  Hamiltonian is thus defined as
\be
H~=H^0 ~+ H^{e-v} ~,
\label{2.1}
\ee
where,
\be
H^0=  \hbar \omega \sum_M \left
(b^\dagger_{ M } b_{M}+\half\right)~+(\epsilon-\mu)\sum_{ms}
c^\dagger_{ms} c_{ms} ~.
\label{2.2}
\ee
$b^\dagger_{M}$ creates a vibron with azimuthal quantum number $M$,  and
$c^\dagger_{ms}$ creates an electron of spin $s$ in an orbital $Y_{1m}$.
By setting $\mu\to \epsilon$ we can discard the second term.

The $H_g$  vibration field is
\be
 u  (\Oh) =   {1\over \sqrt{2}} (Y^*_{2M}(\Oh) b^\dagger_{ M }+Y_{2M}(\Oh)
b_{M}) ~,
\label{2.3}
\ee
where $\Oh$
is a unit vector on the sphere.  The $t_{1u}$ electron field is
 \be
\psi_s(\Oh)=\sum_{m=-1}^1 Y_{1m}(\Oh) c_{ms} ~.
\label{2.4}
\ee
The  electron--vibron interaction  is local and rotationally invariant.
Its form is
completely determined (up to an overall coupling constant $g$) by
symmetry:
\be
H^{e-v}~\propto   g \int\! d\Oh u (\Oh) \sum_s
\psi^\dagger_s(\Oh)\psi_s(\Oh) ~.
\label{2.5}
\ee
Using the relation
\be
\int \!d\Oh ~Y_{LM}(\Oh ) Y_{lm_1}(\Oh) Y_{lm_2}(\Oh) \propto (-1)^M \langle
L,-M|lm_1;lm_2\rangle ~,
\label{2.6}
\ee
where
$\langle \cdots \rangle$ is a Clebsch-Gordan coefficient \cite{ed},
yields the  second quantized Hamiltonian
\bea
H^{e-v}&=&  {\sqrt{3}\over 2} g \hbar \omega \sum_{s,M,m}
(-1)^{m} \left(b^\dagger_M+(-1)^M b_{-M}\right)
\nonumber\\ &&~~~~~~~~~~~~~~~\times  \langle 2, M|1,-m;1,M+m\rangle
c^\dagger_{m  s }c_{M+m s} ~.
\label{2.7}
\eea
The coupling constant $g$ is fixed by the convention of O'Brien, who studied
first this kind of dynamical JT problem \cite{ob1}.  Representation (\ref{2.7})
is convenient for  setting up an exact diagonalization program in the truncated
Fock space.

\subsection{The Real Representation}
The semiclassical expansion  is simpler to derive in the real coordinates
representation. The vibron  coordinates are
 \be
q_{\mu}~={6\over\sqrt{10}}
 \sum_{m=-2}^2 M_{\mu m}  \left(b^\dagger_{m}+(-1)^{m} b_{-m}\right) ~,
\ee
where
\bea
M_{\mu,m \ne 0}&=& (2~ {\rm sign} (\mu) )^{-\frac{1}{2}}
 \left(\delta_{\mu,m} +
 {\rm sign} (\mu)\delta_{\mu,-m}\right),\nonumber\\
 M_{\mu,0}&=& \delta_{\mu,0}.
\label{2.8.0}
\eea
$\{q_\mu\}$ are coefficients of the real spherical functions
\bea
f_\mu(\Oh) &=& {6\over\sqrt{5}} \sum_m M_{\mu,m} Y_{2 m}(\Oh)\nonumber\\
&=& \cases{{6\over\sqrt{10}}~Re\left(Y_{2|\mu|}(\Oh)\right)  & $\mu = 1,2$\cr
 {6\over\sqrt{5}}~ Y_{20}(\Oh)  & $\mu = 0$\cr
{6\over\sqrt{10}}~Im \left(Y_{2|\mu|}(\Oh)\right) & $\mu =-1,-2$}~.
\label{2.8}
\eea
We also choose a real representation for the  electrons
\bea
c^\dagger_{xs}&=& {1\over \sqrt{2}}\left( c^\dagger_{1s} +
c^\dagger_{-1s}\right)\nonumber\\
c^\dagger_{ys}&=& {1\over i\sqrt{2}}\left( c^\dagger_{1s} -
c^\dagger_{-1s}\right)\nonumber\\
c^\dagger_{zs}&=&   c^\dagger_{0s} ~.
\eea
Thus the Hamiltonian in the real representation is given by
\bea
H &=& H^0~+H^{e-v}\nonumber \\
H^0 &=&{ \hbar \omega\over 2} \sum_{\mu }\left( -
\partial_\mu^2   +   q_{\mu}^2\right) \nonumber \\
H^{e-v} &=& g {\hbar\omega\over 2}  \sum_s
(c^\dagger_{xs}, c^\dagger_{ys},c^\dagger_{zs})
\pmatrix{
q_0+\sqrt{3}q_2	&-\sqrt{3}q_{-2}	&\sqrt{3}q_1 \cr
-\sqrt{3}q_{-2}	&q_{0}-\sqrt{3}q_{2}	&-\sqrt{3}q_{-1}\cr
-\sqrt{3}q_1	& \sqrt{3}q_{-1}	& -2q_{0}     }
					\pmatrix{	c_{xs}\cr
                        	                	c_{ys}\cr
							c_{zs} }\nonumber \\
\label{2.9}
\eea

This form of the JT hamiltonian is well known \cite{ob1,vzr}. Since the
Hamiltonian is rotationally invariant, its eigenvalues  are invariant under
simultaneous O(3) rotations of the electronic and vibronic representations.

\section{Jahn--Teller Distortions (Classical)}
\label{class}
In the classical limit, one  can ignore  the vibron derivative terms in
(\ref{2.9}), and treat  $\bq=\{q_\mu\}$ as frozen coordinates in $H^{e-v}$. The
coupling matrix in $H^{e-v}$ is diagonalized by \cite{ob2}:
\be
  T^{-1}(\varpi)~ \pmatrix{z - \sqrt{3}r  &0&0\cr 0& z +
\sqrt{3}r   &0\cr 0&0&-2z } ~T(\varpi) ~,
\label{3.1}
\ee
where
\be
T~=   \pmatrix{ \cos \psi
& \sin \psi& 0\cr -\sin \psi& \cos \psi &0\cr
0&0&1}
  \pmatrix{ \cos\theta &0& \sin\theta\cr
0&1&0\cr
\sin\theta&0&\cos\theta}
  \pmatrix{ \cos\phi &  \sin\phi&0\cr
-\sin\phi& \cos\phi & 0\cr
0&0&1}.
\label{3.2}
\ee
$\varpi=(\phi,\theta,\psi)$  are the three Euler angles of the O(3) rotation
matrix $T$. In the diagonal basis of (\ref{3.1}),  the electron energies depend
only on  two  vibron coordinates:
\be
{\bq}(0)  =\pmatrix{r\cr
                          0\cr
		               z \cr
                          0\cr
                          0} .
\label{3.3}
\ee
By rotating
the vibron coordinates $\bq$ to the diagonal basis using  the  $L=2$
rotation  matrix $D^{(2)}$ \cite{ed},  one obtains
\be
{\bq}_\mu (r,z,\varpi)~= \sum_{m,m',\mu'=-2}^2 M_{\mu,m}
D^{(2)}_{m,m'}(\varpi) M_{m'\mu'}^{-1} {\bq}_{\mu'}(0) ~,
\label{3.3.1}
\ee
where $M_{\mu,m}$ was defined in (\ref{2.8.0}).

By (\ref{3.3.1}), and the unitarity of $D$ and $M$, $|{\bq}|^2$ is invariant
under rotations of $\varpi$. Thus, the adiabatic potential energy $V$ depends
only on $r$,  $z$, and the occupation numbers of  the electronic eigenstates $
n_i$, where  $\sum_i n_i=n$.
\be
V(z, r,[n_i]) = {\hbar\omega\over 2} ( z^2+r^2)~+  {\hbar\omega
g\over 2}\left( n_1(z-\sqrt{3}r) +n_2(z+\sqrt{3}r) - n_3 2 z \right)  .
\label{3.4}
\ee
$V$ is minimized at the JT distortions
$({\bar z}_n,{\bar
r}_n,{\bar n}_i)$, at which the classical energy is given by
\be
E_n^{cl} ~= \mbox{min}~ V({\bar z}_n,{\bar
r}_n,{\bar n}_i) .
\label{3.5}
\ee
The JT distortions at different fillings are given in Table I. We define
${\tilde \phi},{\tilde\theta}$ as the longitude and latitude with respect to
the diagonal frame (``principal axes'') labelled $(1,2,3)$ ($3$ is at the north
pole). ${\bar z},{\bar r}$ parametrize the Jahn-Teller distortion in the real
representation (\ref{2.8}), as
\be
\langle u^{JT}({\tilde\theta},{\tilde\phi})\rangle ~=
 { {\bar z}\over 2}(3\cos^2{\tilde\theta} -1 )~+
 {{\bar r} \sqrt{3}\over 2} \sin^2{\tilde\theta}\cos(2{\tilde\phi}).
\label{3.6}
\ee
In Table I we present the values of the ground state JT distortions at
all electron fillings. We see that electron fillings $n$ = 1, 2, 4, 5 have
{\em unimodal distortions } which are symmetric about the $3$ axis,  while $n=3
$ has a {\em bimodal}, about the $3$ and $1$ axes. The two types of
distortions are portrayed in Fig. \ref{dist}. we depict the distortions of
(\ref{3.6}) for the unimodal and bimodal cases.

\section{Semiclassical Quantization}
\label{semiclas}
At finite coupling constant $g$,  quantum fluctuations about the frozen JT
distortion must be included. In order to carry out the semiclassical
quantization, we define a  natural set of five dimensional coordinates
$r,z,\varpi$.  $\varpi$ parametrize the motion in the JT manifold (the valley
in the ``mexican hat'' potential $V$) and $r,z$ are  transverse to the JT
manifold, since $V$ depends on them explicitly.   The  transformation ${\vec
q}(r,z,\varpi)$  was given in (\ref{3.3.1}), and was derived explicitly in
Ref. \cite{ob2} to be
\bea
q_2 &=&  z  \half\sqrt{3} \sin^2 \theta \cos2\phi  + r\half
(1+\cos^2\theta)\cos 2\phi \cos 2\psi\nonumber\\
&&~~~~~~~~~~~-r\cos\theta\sin 2\phi \sin 2\psi\nonumber\\
q_1 &=&  z  \half\sqrt{3} \sin 2\theta \cos \phi  - r\half
\sin 2\theta \cos \phi \cos 2\psi\nonumber\\
&&~~~~~~~~~~~+  r \sin\theta\sin \phi \sin 2\psi\nonumber\\
q_0 &=& z \half (3 \cos^2\theta -1) + {\bar r}\half
\sqrt{3}\sin^2\theta \cos 2\psi\nonumber\\
q_{-1} &=& z \half\sqrt{3} \sin 2\theta \sin \phi  - r\half
\sin 2\theta \sin \phi \cos 2\psi\nonumber\\
&&~~~~~~~~~~~-r\sin\theta\cos \phi \sin 2\psi\nonumber\\
q_{-2} &=& z \half\sqrt{3} \sin^2 \theta \sin 2\phi  + r\half
(1+\cos^2\theta)\sin 2\phi \cos 2\psi\nonumber\\
&&~~~~~~~~~~~-r\cos\theta\cos 2\phi \sin 2\psi ~.
\label{4.2}
\eea

The velocity in $R^5$ is given by
\be
{\dot\bq}(r(t),z(t),\varpi(t))=\partial_r\bq {\dot r} + \partial_z\bq {\dot z}+
\partial_\varpi \bq \cdot {\dot\varpi} ~.
\label{4.3}
\ee
Using (\ref{4.2}) and (\ref{4.3}), we calculate the classical kinetic
energy in terms of the JT coordinates. After some cumbersome, but
straightforward, algebra  the kinetic energy is obtained in the compact and
instructive form:

\bea \frac{1}{2} |{\dot {\vec q}}|^2&=& \frac{1}{2}\left( {\dot z}^2 + {\dot
r}^2 +  \sum_{i=1}^3 I_i \omega_i^2 \right),\nonumber\\
\omega_1 &=& -\sin \psi {\dot\theta} + \cos\psi\sin\theta {\dot \phi} ,
\nonumber\\
\omega_2 &=& \cos \psi {\dot\theta} + \sin\psi\sin\theta {\dot
\phi},\nonumber\\
\omega_3&=&{\dot \psi} +\cos\theta{\dot\phi} \nonumber\\
(I_1,I_2,I_3)&=& \left((\sqrt{3}  z+ r)^2, (\sqrt{3}  z- r)^2 ,4 r^2\right).
\label{4.4}
\eea
For finite JT distortions, we can identify  the Euler angles terms as the
kinetic energy of a rigid body rotator \cite{com}, and the quantities
$I_i({\bar z},{\bar r})$ as moments of inertia in the principle axes frame
1,2,3. Thus, the
Euler angles dynamics follows that of a {\em rigid body rotator} \cite{ed}.

The unimodal and bimodal cases will be discussed separately.

\subsection{Unimodal Distortions}
\label{unimodal}
For the unimodal cases (which we found for the ground states of $n=1,2,4,5$,
${\bar r} = 0$ on the JT manifold. The ``moments of inertia''  in (\ref{4.4})
are given by the tensor
\be
 {\hat I} = 3{\bar z}^2 \pmatrix{1&0&0\cr
                              0&1&0\cr
                               0&0&0} .
\label{4.5}
\ee
This corresponds to the rotational energy of a point particle on a sphere,
which is described by the angles $\theta,\phi$, and moment of inertia $3{\bar
z}^2$. Since axis $3$ has no ``mass'', its angular velocity is dominated by
$\dot\psi$. This implies that we must keep the term  $r^2{\dot\psi}^2$ but can
discard the smaller mixed terms ${\dot\psi}{\dot\phi}$. This yields
 \be
\frac{1}{2} |{\dot {\vec q}}|^2 \approx  \half \left( {\dot z}^2 + {\dot r}^2
+  r^2 (2{\dot \psi})^2 ~+  3{\bar z}^2\left( {\dot\theta}^2 + \sin^2\theta
{\dot \phi}^2 \right) \right) .
\label{4.6}
\ee
The angular velocity ${\dot\psi}$ couples to $r^2$ as
in the kinetic energy of a three dimensional vector $\br$  parameterized by the
cylindrical coordinates
\be
\br ~= (r\cos(2\psi ),r\sin(2\psi ),z-{\bar z}).
\label{4.7}
\ee
For  $|\br|<< {\bar z}$, the potential is simply
\be
V(\br) ~\approx \half |\br|^2 ~.
\ee
Thus, the semiclassical Hamiltonian of the unimodal distortion is
\bea
H^{uni} &\approx& H^{rot} + H^{ho}\nonumber\\
H^{rot} &=& {\hbar \omega \over 6{\bar z}^2} {\vec L}^2\nonumber\\
H^{ho}&=&{\hbar \omega} \sum_{\gamma=1}^3 (a^\dagger_\gamma a_\gamma + \half)
{}~,
\label{4.10}
\eea
where  ${\vec L}$ is an angular momentum operator, and $H^{ho}$ are the three
harmonic oscillator modes of $\br$.
The energies are given by
\be
E^{uni} ~= {\hbar \omega }\left({1\over 6{\bar z}_n^2 }L(L+1) +
 \sum_{\gamma=1}^3
(n_\gamma +\frac{1}{2} )\right) .
\label{4.10.1}
\ee
The rotational part of the eigenfunctions is
\be
\Psi^{rot}_{Lm} ({\bq}) =  Y_{Lm}(\Oh) ~  |[n_{is}]\rangle_\Oh ~,
\label{4.11}
\ee
where $\Oh=(\theta,\phi)$ is a unit vector, and $ |[n_{is}]\rangle_\Oh$ is the
electronic adiabatic ground state. It is a Fock state in the {\em principal
axes}  basis.  In terms of the stationary Fock basis $|[n_{\alpha s'}]\rangle$
where $\alpha=x,y,z$, the adiabatic ground state is
\be
|[n_{is}]\rangle_\Oh ~= \sum_{[n_{\alpha s }]}
   \langle[n_{\alpha s }] |[n_{is}]\rangle_\Oh
 |[n_{\alpha s }]\rangle .
\label{4.12}
\ee
Each overlap is a Slater determinant which is a sum of $n$ products of
spherical harmonics
\be
\langle[n_{\alpha s}] |[n_{is}]\rangle_\Oh ~=
\sum_{[\nu]} C_{[\nu]} Y_{1\nu_1}(\Oh) Y_{1\nu_2}(\Oh)\cdots Y_{1 \nu_n}(\Oh)
{}~,
\label{4.12.1}
\ee
where $C_{[\nu]}$ are constants.

Now we discuss how boundary conditions determine the allowed values of $L$.
A reflection on the JT manifold  is given by
\be
\Oh ~\to -\Oh .
\label{4.13}
\ee
Spherical harmonics are known to transform under reflection as
\be
Y_{L m} \to\!(-1)^LY_{L m} ~.
\ee
Thus, by (\ref{4.12}) and (\ref{4.12.1}), the electronic part of the wave
function transforms as
\be
|[n_{is}]\rangle_\Oh \to (-1)^n |[n_{is}]\rangle_{ -\Oh } ~.
\label{4.14}
\ee

The reflection (\ref{4.13}) can be performed by moving on a {\em  continuous
path} on the sphere from any point to its opposite. (See Fig. \ref{uniorbit}).
It is easy to verify, using (\ref{3.3.1})  or (\ref{4.2}), that this path {\em
is a closed orbit of ${\vec  q} \in R^5$}:
\be
{\bq}(\Oh)\to {\bq}(-\Oh) ={\bq}(\Oh) ~.
\label{4.15}
\ee
Thus we find that the electronic wave function yields a {\em Berry phase
factor} of $(-1)^n$ for rotations between opposite points on the sphere  which
correspond to closed orbits of ${\bq}$. In order to satisfy (\ref{4.11}))
using  the invariance  of the left hand side under reflection, the
pseudorotational $Y_{Lm}$ wavefunction must cancel the electronic Berry phase.
This amounts to a {\em selection rule} on $L$:
\be
(-1)^{L+n} = 1 ~.
\label{4.16}
\ee
Thus, the ground state for $n = 1$ and $5$ electrons has pseudo-angular
momentum
$L = 1$ and finite zero point  energy due to the non trivial Berry phases.

\subsection{Bimodal Distortion}
The analysis of the bimodal distortions  $n = 3$ proceeds along similar lines.
The distortion obeys
\be
{\bar z}= \sqrt{3}{\bar r} ~.
\ee
{}From Eq.(\ref{4.4})  we see that the kinetic energy is given by
\be
\frac{1}{2} |{\dot {\vec q}}|^2 =
 \frac{1}{2}\left( {\dot z}^2 + {\dot r}^2 + \sum_{i=1}^3 I_i
\omega_i^2\right),
\label{4.20.0}
\ee
where the  inertia tensor is
\be
 {\hat I} = 2{\bar z}^2 \pmatrix{4&0&0\cr
                              0&1&0\cr
                               0&0&1} .
\label{4.20}
\ee
 The quantization of the pseudo--rotational part
 is the quantum symmetric top Hamiltonian. Fortunately, it is
a well-known textbook problem (see e.g. Ref. \cite{ll,ed}).
The eigenfunctions of a rigid body rotator are the rotational matrices
\be
D_{mk}^{(L)} (\varpi) ,
\ee
where $ L,m,k $ are quantum numbers of the commuting operators ${\vec L}^2,
L^z, L^1$ respectively. $L^z$ and $L^1$ are defined with respect to the fixed
$z$ axis and the co-rotating $1$ axis respectively. The quantum  numbers are in
the ranges
\bea
L&=& 0,1,\ldots \infty\nonumber\\
m,k &=& -L, -L+1,\ldots L ~.
\label{4.20.1}
\eea
The remaining coordinates are two massive
harmonic oscillators modes
\be
\br = (r-{\bar r},z-{\bar z}).
\label{4.21}
\ee
The semiclassical Hamiltonian is thus
\bea
H^{bi} &\approx& H^{rot} + H^{ho},\nonumber\\
H^{rot} &=& {\hbar \omega \over 4{\bar z}^2 } {\vec L}^2 -
{3\hbar \omega \over 16{\bar z}^2} (L^1)^2,\nonumber\\
H^{ho}&=&{\hbar \omega} \sum_{\gamma=1}^2 (a^\dagger_\gamma a_\gamma + \half),
\label{4.22}
\eea
and its eigenvalues are
\be
E^{bi}={\hbar \omega }\left({1\over 4{\bar z}^2  }L(L+1)-
 {3\over 16{\bar z}^2}k^2~+\sum_{\gamma=1}^2
(n_\gamma +\frac{1}{2} )\right) .
\label{4.23}
\ee
The rotational eigenfunctions are explicitly dependent on $\varpi$ as
\be
\Psi^{rot}_{Lmk}[{\bq}] =   D_{mk}^{(L)} (\varpi)
\prod_{is}|n_{is}\rangle_{\varpi} ~ .
\label{8.1}
\ee

\subsection{Berry Phases of a Bimodal Distortion}

Unlike the unimodal case, in the bimodal case no single reflection fully
classifies the symmetry of the wavefunction.  However, one can obtain definite
sign factors by transporting the electronic ground state  in certain orbits. We
define the rotations of $\pi$ about principle axis $L^i$  as $C_i$, which are
schematically depicted in Fig. \ref{biorbit}. The Berry phases associated with
these rotations can be read directly from the rotation matrix $T$ in  Eq.
(\ref{3.2}).  For example: for $\psi\!\to\!\psi+\pi$ ($C_3$), the states
$|1\rangle$ and $|2\rangle$ get  multiplied by $(-1)$.

Since $D^{(L)}_{m,k}$ transform  as $Y_{Lk}$ under $C_i$,  it is easy to
determine the  sign factors of the pseudorotational wavefunction. The results
are given below:
\begin{eqnarray}
C_1:|1,0,2\rangle_{\varpi}  \to |1,0,2\rangle_{\varpi'} &~~~& C_1:D^{(L)}_{m,k}
\to (-1)^k D^{(L)}_{m,k} \nonumber\\
C_2:|1,0,2\rangle_{\varpi}  \to -|1,0,2\rangle_{\varpi'} &~~~&
C_2:D^{(L)}_{m,k}
\to (-1)^{L+k} D^{(L)}_{m,-k}
\nonumber\\
C_3:|1,0,2\rangle_{\varpi}  \to -|1,0,2\rangle_{\varpi'} &~~~&
C_3:D^{(L)}_{m,k}
\to  (-1)^{L} D^{(L)}_{m,-k} ~.
\label{9}
\end{eqnarray}
${\bq}$ are coefficients in an $L=2$ representation, and therefore are
invariant under $C_1,C_2,C_3$. $C_i$ describe continuous closed orbits in
$R^5$.
In order to satisfy (\ref{9}) and using the degeneracy of $E^{bi}$ for
$k\!\to\!-k$,  we find that
\be
L~=~\mbox{odd}~,~~~~~k ~=~\mbox{even}.
\ee
In
particular, the ground state
of (\ref{8.1}) is given by
$L\!=\!1$, and $k\!=\!0$.

\subsection{High-Spin Polarized Ground States}

It is possible to repeat the semiclassical analysis assuming that the spins are
maximally polarized. These high-spin states are important, as they tend to
prevail for strongly repulsive intra-level Hubbard $U$ (Hund's rule)
situations. In this case, we determine the JT distortions considering the Pauli
exclusion between likewise spins.  In Table II the JT distortions of the spin
polarized ground states are listed. Our results for $n$=2,4 ($S$=1), and $n$=3
($S$=3/2) cases are presented. The latter is trivial, since in that case
$n_1=n_2=n_3=1$, and therefore there is no JT effect at all. For $n$=2, ($S$=1)
there is unimodal distortion of  ${\bar z}=-g$  which is smaller than the
unpolarized ground state, and is equal to the distortion of the $n=5$ case.
Inspection of the orbital energies $\epsilon_1 = \epsilon_2 = -g ~,~ \epsilon_3
= 2g$ provides a clear explanation for the identical distortions of the $n$=2
($S$=1) and $n$=5 ($S=\half$) cases, since in both cases $\epsilon_3$ is
occupied by a ''spin up hole``.

Electronically, however, the two states are very different. First, we do not
have a Berry phase for even number of electrons, as the individual
contributions from each of the two electrons cancel  out. Second, there is a
nonzero {\em electronic} orbital angular momentum. For example, the symmetry of
the two-electron state prior to JT distortion is $^3$P (i.e. $^3$$t_{1u}$), and
so it remains following dynamical JT \cite{ham}. At finite coupling the two
electrons in their ground state are still coupled in a $^3$P electronic state,
with $L_{orb}$=1, where $L_{orb}$ is the electronic orbital angular momentum,
not to be confused with  the pseudorotational quantum number $L$.  Due to the
absence of a Berry phase $L$ must in fact be even, in contrast with the single
electron case, and in agreement with Eq. (\ref{4.16}). Thus although both cases
have  threefold degeneracies, they arise from different physical motion: purely
electronic (for the $n$=2, $S$=1 case) versus mixed electron--vibron motion (in
the $n$=5, $S$=1/2 case).

\section{Exact Diagonalization}
\label{exact}

The above semiclassical scheme gives a clear and intuitive picture of the
behaviour of the system in a strong coupling limit. This limit is appropriate
for describing, e.g., Na$_3$ \cite{delac}. However, in C$_{60}$ the actual
range of the coupling parameter - $g\approx 0.3$ for a typical mode
\cite{antr,II} - suggests that the electron--vibron coupling is actually in the
weak to intermediate regime.

Here we diagonalize the electron--vibron Hamiltonian (\ref{2.7}) for single
$H_g$ mode  in a truncated Fock space. This approach yields accurate results
unless the coupling strength is too large, and the higher excited vibrons admix
strongly into the low lying states.  We compare the results to the asymptotic
large $g$ expressions of the semiclassical approximation. The ground state
energy for $n=1$ has  been previously computed in this fashion by O'Brien
\cite{ob2}. Here we present detailed results for all electron occupations, and
also for the excitation spectra.

Our basis is the finite dimensional  Fock space  of electrons
and vibrons, \be
\left\{ ~ |
n_{ M },n_{ms} \rangle~~: ~~ N_v  \le N^{max},~\sum_{ms} n_{ms}=n ~ \right\} ,
\label{4.1}
\ee
where  $N_v=\sum_M n_{ M }$ is the total vibron occupation. By gradually
increasing $N^{max}$, we have found empirically that accurate results can be
obtained for $g \le N^{max}/2$, for  levels with unperturbed energy below
$\hbar\omega N^{max}/2$.  In particular, we have chosen $N^{max}=5$ (for
$n=2,3$)  which yields an accuracy of better than $0.05\hbar \omega$ for $g\leq
0.6$ and levels with $N_v \leq 1$. The effect of truncation is a general upward
shift of the levels, which gradually  increases for higher excited levels.
Level splittings and excitation energies are therefore  less sensitive to the
cutoff error.

In Figures \ref{ExactS1}, \ref{ExactS2} and \ref{ExactS3} the ground state and
a  few of the excited states energies are plotted for one two and three
electrons respectively. The four and five electron spectra are related to the
two and one electron spectra by particle--hole symmetry. Energies are plotted
as functions of $ g^2 $. We compare the results to the semiclassical
expressions (\ref{4.10.1}) and  (\ref{4.23}) for large coupling, and to second
order perturbation theory at weak coupling. We discuss the different cases in
detail, below.

\subsection{$n=1,5$  electrons}

The ground state for one electron or hole in the $t_{1u}$ shell is a
threefold-degenerate state (all degeneracies given do not include spin) of the
same symmetry: this fact is in complete analogy with what happens in the $e
\otimes E$ coupled system, where the final dynamical JT coupled ground state
has again $E$ symmetry \cite{ham}. Additional splitting of this ground state
could occur via spin-orbit coupling, not included in the present treatment.
Recent spectroscopic data \cite{gasyna} of C$_{60}^-$ embedded in solid Ar
confirm indirectly the presence of the pseudorotational $L=1$ ground state
degeneracy, through direct observation of spin-orbit splittings of about 30 and
of 75 cm$^{-1}$ for the $t_{1u}$ ground state and for the $t_{1g}$ excited
electronic state $\approx$ 1 eV above. The decrease of ground state energy is
initially fast, and becomes gradually slower for increasing $g$. We shall
return to this point in detail in \cite{II}.

As shown in Fig. \ref{ExactS1}, for large $g$, the $n$=1 ground state energy
correctly approaches the strong coupling limit
\be
E~\sim -\half g^2+\frac{3}{2}+\frac{1}{3g^2} ~,
\ee
except for a small shift due, as mentioned above, to a finite-cutoff error.
Above the ground state, there are families of excitations, corresponding to
increasing values of $N_v$. The lowest, for $N_v=1$, comprises $3\times5=15$
states, since for $n$=1, $N_v$=1 there are just 3 electron states and 5 vibron
states available. These states correspond to a direct product of a P
(electronic) and a D (vibrational) manifold. As elementary angular momentum
theory requires, they split into $L$ = 3, 2 and 1 levels, which are found, in
order of increasing energy. The splitting initially is proportional in $g^2$,
for small $g$, with significant deviations from linearity at $g^2\approx 0.2$.
As coupling increases, we note the slower downward trend of the {\em even} $L$
states, than both the ground state and the associated ''soft`` odd-$L$
excitations. This clearly reflects the Berry phase selection rule (\ref{4.16})
that {\em no even $L$ should appear among the low lying excited states} in
strong coupling. The lowest excitation from the ground state is $L$=1$ \to
L$=3, anticipating already at very weak coupling the strong coupling result
that this excitation energy should fall fastest, and collapse as
$\frac{5}{3g^2}$. Unlike the $L$=3 state, the $L$=2 and $L$=1 excited states do
not show any tendency to collapse onto the ground state in the large $g$ limit.
Therefore they can be seen as modes involving essentially radial massive
vibrations.

The next group of excitations is for $N_v=2$, and comprise $3\times15=45$
states. This multiplet splits into seven levels corresponding to  $L$ = 5, 3,
1, 4, 2, 1 and 7. The lowest ($L$=5) level crosses two levels of the lower
($N_v$=1) multiplet in its downwards motion to become the second excited state
above the $L$=3 level, eventually constituting the low energy odd-$L$
rotational multiplet of the strong coupling picture. The same route is followed
by the lowest level of $N_v=3$, which is an $L$=7 state. In fact, all the
lowest split levels from each $N_v$ multiplet appear to have $L$=$2N_v+1$ and
follow the same route.

For $N_v=2$ we can similarly follow the movement with $g$ of the $L$=4 level
which decreases slowly towards the $L$=2 state from the lower $N_v=1$ to
add to the group of massive radial vibrations.

\subsection{ $n=2,4$ electrons}

Figure \ref{ExactS2}  has several features which contrast sharply with the one
electron case. The $N_v=0$ multiplet, has 15 two-electron  states. The spin
singlet subspace constitutes of a 6-fold degenerate multiplet that splits into
an orbital S and a D multiplet. As the semiclassical Eq. (\ref{4.16}) suggests,
the ground state and lowest excitations in the strong coupling limit have
orbital degeneracies of {\em even} angular momenta. In fact, the lowest two
among these states ($L$ = 0, 2) both come from the $N_v$=0 multiplet, in
contrast with the one electron case. The next pseudorotational level ($L=4$)
originates in the $6\times5=30$-fold degenerate $N_v=1$, spin singlet
multiplet. Actually, at weak coupling it starts out being second in the
ordering ($L$= 2, 4, 3, 2, 1, 0), but already at very small $g$ it crosses the
lower L=2 partner and approaches the pseudo--rotational asymptotic level. The
convergence with increasing cutoff $N^{max}$ is worse than in the $n$=1 case,
which can be as due to larger JT distortions associated with  two electrons.
The spin triplet ($S$=1) states of $n$=2 have not been plotted, as they behave
in exactly the same fashion as the $n$=1 states (see Fig. \ref{ExactS1}). This
figure can be read in terms of $n$=2 $S$=1 states simply by replacing the spin
multiplicity label 2, as was in the case $n$=1, with 3. By comparison of Fig.
\ref{ExactS2} and Fig. \ref{ExactS1} we notice that the low-spin $^1$P state
of $N_v$=1 is exactly degenerate with the high-spin $^3$D state in the same
multiplet. This degeneracy seems accidental.

\subsection{$n=3$ electrons}

For three electrons, the results are shown in Fig. \ref{ExactS3}. The 8-fold
degenerate $N_v=0$ multiplet splits into two states characterized by
degeneracies 3 and 5 ($^2$P and $^2$D). The ground state has the correct
symmetry for an $L$=1, $k$=0 state, which is predicted to be the ground state
in the semiclassical limit. We also expect the lowest excitations to be
classified as $L$=3,~$k$=2 (14-fold degenerate), and $L$=3,~$k$=0 (7-fold
degenerate). In fact, three levels from the $N_v=1$ multiplet move down toward
the ground state for increasing $g$. The one which moves lowest is 9-fold
($^2$G). In the $g \to \infty$ limit, it must therefore merge with the 5-fold
levels from the $N_v$=0 multiplet to produce the expected $L$=3,~$k$=2
pseudo--rotator excitation. The next excitation of $L$=3,~$k$=0 state can be
identified as an asymptotic limit of the $^2$F 7-fold degenerate state seen
in Fig. \ref{ExactS3}.

A remarkable feature of the $n=3$ case is the presence in the $N_v=1$ multiplet
of a state (the $^2$S) whose energy is independent of $g$!  This
state is degenerate with the  $S$=3/2 state $^4$D which has no
JT distortion.

\section{Pair Binding Energies}
\label{pair-bind}
The {\em pair energy} for an average filling of $n$ electrons is defined as
\be
U_n= E_{n+1}+E_{n-1} -2E_n   ~,
\label{5.1}
\ee
where $E_n$ are the fully relaxed ground state energies of  $n$ electrons.
Formally, $U$ is the real part of the two--electron vertex function at zero
frequency. If this energy is negative for odd values of $n$, it means that
electrons will have lower total energy if they separate into $(n-1)$ and
$(n+1)$ occupations of different molecules, rather than occupying $n$ electrons
on all molecules. For odd values of $n$, this is an effective pairing
interaction often called ``pair binding'' in the literature \cite{ck}.
In Section \ref{semiclas} we found that for all odd $n$, the pair energies are
negative, and given by the large $g$ asymptotic expression
\be
U_{n=1,3,5}~\sim - g^2 + 1 -
{2\over 3 g^2} ~+{\cal O}(g^{-4}).
 \label{5.2} \ee
The first term is  the {\em classical energy}. The second term is due to
reduction of zero point energy along the JT manifold, since only
radial modes remain hard. This term is independent of $g$ and positive. The
last term is due to the quantum pseudo--rotator Hamiltonian, and the Berry
phases which impose a finite ground state energy associated with odd $L$ for
odd numbers of electrons. This term, although nominally small at large $g$,
becomes important at weaker coupling. If (\ref{5.2}) is extrapolated to the
weak coupling regime the last term would dominate the pair binding energy. The
exact diagonalization shown in Figure \ref{ExactU},  indeed shows a significant
enhancement of the pair binding energy over the classical value in the weak
coupling regime.

In the weak coupling limit, we can obtain analytical expressions for $U_n(g)$
for $g<<1$ by second order perturbation theory.  The unperturbed Hamiltonian is
the non interacting part  $H^0$  with eigenstates (\ref{4.1}). The perturbing
hamiltonian is $H^{e-v}$ of Eq. (\ref{2.7}), which connects Fock states
differing by one vibron occupation. All diagonal matrix elements vanish, and
the leading order corrections to any degenerate multiplet are of order $g^2$.
These are given by diagonalization of the matrix \cite{sak},
\be
\Delta^{(2)}_{n_{ms},n_{ms}'}=
 \langle 0, n_{ms} | H^{e-v} {1 \over {E^{(0)}_a-H^0}}
H^{e-v} | 0,n_{ms}' \rangle ,
\label{5.3}
\ee

in the degenerate 0-vibrons subspace. The sum implied by the inverse operator
$(E^{(0)}_a-H^0)^{-1}$ extends just to the $N_v$=1 states. The eigenvalues of
$\Delta^{(2)}$  yield the ground state energies and splittings for different
electron fillings.  These results, for all $H_g$ and also $A_g$ modes, and
extended to the $N_v$=1 multiplet, will be discussed more extensively in
\cite{II}.

Here we refer only to ground state energetics. In particular, using the
perturbative expressions, we obtain, for a single $H_g$, mode the small $g$
pair binding energy
\be
{ U_{n=1,3,5}\over \hbar\omega}~= -{5\over 2} g^2~+ {\cal O}(g^4).
\label{5.6}
\ee
The dependence strictly on powers of $g^2$ alone, with absence of all odd
powers, is a consequence of the already mentioned
$\Delta N_v = \pm 1$ selection rule of Eq. (\ref{2.7}).
The origin of the 5/2 factor that characterizes the perturbative
result (\ref{5.6}) with respect to the classical pair binding energy (Table I)
will be discussed in Ref. \cite{II}.

The molecular pair binding
energy  can be considered as an effective negative-U Hubbard interaction for
the
lattice problem, provided that the Fermi energy $\epsilon_F$ is not much larger
than the JT  frequency scale $\omega$.
A mean field estimate of the transition temperature for the negative-$U$
Hubbard
model in the weak coupling regime is \cite{NegU,ck}
\be
T_c \approx \epsilon_F
\exp \left[ \left(- N(\epsilon_F) | U |\right)^{-1} \right] .
\label{5.7}
\ee
In Refs. \cite{sch} and \cite{lannoo}, the results of Migdal--Eliashberg
approximation for the superconducting transition temperature was given. Without
the Coulomb pseudopotentials  this approach yields
\bea
T_c &\approx& \omega \exp \left[ \left(- N(\epsilon_F) |V| \right)^{-1} \right]
 \nonumber\\
V&=&-{5\over 6} g^2 ~ .
\label{5.8}
\eea
By comparing (\ref{5.6}) to (\ref{5.8}) we find a striking discrepancy between
the values of the effective pairing interaction:
\be
U~=3 V .
\ee
That is to say: in  the weak coupling regime, the correct molecular calculation
yields  a pairing interaction which is {\em three times larger} than the
results of Migdal--Eliashberg theory.

\section{Discussion}
\label{discuss}
In this paper, we have solved the problem of a single $H_g$ vibron
coupled to $t_{1u}$ electrons in a C$_{60}^{n-}$ molecule. The model
is too simplified for quantitative predictions for C$_{60}$, but it contains
interesting  novel physics which will be important for further
studies of this system.

Semiclassically, a dynamical Jahn--Teller effect occurs.  For $n=1,2,4,5$, the
molecule distorts unimodally, giving rise to a pseudo-angular momentum
spectrum, plus three harmonic oscillators. For $n\!=\!3$, there is a bimodal
distortion, which generates a spectrum of a symmetric top rotator, plus two
harmonic oscillators. The pseudo rotations are subject to non trivial Berry
phase effects, which determine the  pseudo-angular momenta $L$, and thus the
degeneracies and level ordering of the low lying states. Strong Berry phase
effects seem to survive even at moderate and weak coupling as shown by the
exact diagonalization results.

We find at weak coupling that the pair binding energy is a factor of $5/2$
larger than the classical JT effect, and a factor of three larger
than the pairing interaction of  Migdal--Eliashberg theory of
superconductivity. This enhancement can be interpreted semiclassically as due
to large zero point energy reduction of the pseudo--rotations. From the weak
coupling point of view, this effect is due to degeneracies in both electronic
and vibronic systems.

Migdal's approximation neglects vertex corrections in the resummation of
two--particle ladder diagrams. This is justified only in the retarded limit
$\omega << \epsilon_F$. Here we have considered the  opposite limit, where the
molecular ground state energies are solved first, assuming that the JT
relaxation time is  of the same order, or faster than the inter molecular
hopping time. In this regime, we have found therefore that Migdal's
approximation  substantially {\em underestimates} the pairing interaction, and
$T_c$, for these ideal molecular solids \cite{pietron}. This large effect
suggests that some of the enhancement is likely to carry over to the real case
of A$_3$C$_{60}$ metals, where electron hopping $t$ and vibron frequencies are
of similar strength.

In Part II we shall consider a more realistic model which includes all
important vibron modes of  C$_{60}$.  We shall present quantitative
predictions for the electron--vibron effects on the spectroscopy of C$_{60}$
ions.

\subsubsection*{Acknowledgements}
A.A. is indebted to Yossi Avron and Mary O'Brien for valuable discussions, and
acknowledges the Sloan  Foundation for a fellowship. ET and NM wish to
acknowledge discussions with S. Doniach. This paper was supported by grants
from the US-Israel Binational Science Foundation,  the Fund for Promotion of
Research at the Technion, and the US Department of Energy  No.
DE-FG02-91ER45441, the Italian Istituto Nazionale di Fisica della Materia INFM,
the European US Army Research Office, and NATO through CRG 920828. \vfill\eject

\vfill\eject

\centerline{\underbar{\bf Table I}}
\vskip 0.3in
\begin{tabular}{|c|c|l|c|c|c|} \hline
$n$&$S$& $({\bar z}_n,{\bar r}_n)$ & $({\bar n}_1,{\bar n}_2,{\bar n}_3)$ &
$ E_n/(\hbar\omega)$ &$U_n/(\hbar \omega)$ \\ \hline
0 &0		& $(0 ,0)$ & (0,0,0)&  ${5\over 2}$&   \\
1 &$\frac{1}{2}$& $(g ,0)$  & (0,0,1)&  $-\half g^2 +{3\over 2}+{1\over
3g^2} $&   $  -g^2+1-{2\over 3g^2}  $
\\ 2 &0		& $(2g ,0)$ & (0,0,2) &
$-2g^2+{3\over 2}  $&  \\
3
&$\frac{1}{2}$& $({3\over 2} g ,  {\sqrt{3}\over 2} g )$  & (1,0,2) &
$ -{3\over 2}
g^2+1+{1\over 3g^2} $& $ -g^2+1-{2\over 3g^2}  $       \\
 4 &0		& $(-2g
,0)$  & (2,2,0)  &   $-2g^2+{3\over 2}  $&  \\
5
&$\frac{1}{2}$& $(-g ,0)$
 & (2,2,1)&  $ -\half g^2 +{3\over 2}+{1\over
3g^2}    $& $   -g^2+1-{2\over 3g^2}     $  \\
6 &0		& $(0 ,0)$ & (2,2,2)& ${5\over 2}$ &  \\ \hline
\end{tabular}
\vskip 0.7in

TABLE I. Semiclassical ground state distortions and energies for a single $H_g$
coupled mode of frequency $\omega$. $n$ is the electron number, $S$ is the
total spin, ${\bar z}_n,{\bar r}_n$ are the JT distortions, ${\bar n}_i$ is the
occupation of orbital $i$, $E_n$ is the ground state energy and, and $U_n$ is
the pair energy (Eq.(\ref{5.1})). Energies are calculated for strong coupling
to order $g^{-2}$.

\vskip 0.7in
\centerline{\underbar{\bf Table II}}
\vskip 0.3in
\begin{tabular}{|c|c|l|c|c| } \hline
$n$&S& $({\bar z}_n,{\bar r}_n)$ & $({\bar n}_1,{\bar n}_2,{\bar n}_3)$ &
$ E_n/(\hbar\omega)$ \\ \hline
2 &1            & $(-g ,0)$  & (1,1,0) &  $-\half g^2+{3\over 2}  $ \\
3 &$\frac{3}{2}$& $(0,0)$       & (1,1,1)       & ${5\over 2}$    \\
4 &1            & $(g ,0)$ (uni) & (1,1,2)&  $-\half g^2 +{3\over 2}  $  \\
\hline
\end{tabular}
\vskip 0.7in

TABLE II. High spin ground state properties, in the same notation of Table I.
\newpage

\begin{figure}
\caption{A polar representation of the Jahn-Teller distortions
$u^{JT}({\tilde \theta},{\tilde \phi})$, Eq.(\protect\ref{3.6}).
The distortion is measured relative to a sphere.
(a) The unimodal distortion for the
ground states of $n=1,2,4,5$ electrons (b) The bimodal distortion
for $n=3$ electrons.}
\label{dist}
\end{figure}

\begin{figure}
\caption{Berry phase calculation for unimodal distortions. An path between
reflected points on the unit  sphere corresponds to a closed orbit in the five
dimensional ${\bq}$-space. According to Eq. (\protect\ref{4.14}), such a path
acquires a Berry phase of $(-1)^{n}$ from the $n$-electron  wavefunction.}
\label{uniorbit}
\end{figure}

\begin{figure}
\caption{Berry phases calculation for the bimodal distortion ($n=3$).  $\varpi$
are the three Euler angles which rotate the principal axes of the bimodal
distortion. $C_i$ denote rotations by $\pi$ around corresponding axes. On the
right we depict the electronic Berry phases associated with the three closed
orbits in ${\vec q}$-space, given by Eq.~(\protect\ref{9}).}
\label{biorbit}
\end{figure}

\begin{figure}
\caption{Exact spectrum for one electron as a function of the square
electron--vibron coupling constant $g^2$.  The vibron occupations are truncated
at $N^{max}=5$. The semiclassical energies (Eq. (\protect\ref{4.10.1})) are
drawn by dashed lines for the lowest three pseudorotational multiplets
($n_\gamma$=0, $L$=1, 3, 5). The unit of energy  is the vibron quantum
$\hbar\omega$.}
\label{ExactS1}
\end{figure}

\begin{figure}
\caption{Exact spectrum for two electrons ($S=0$). The
semiclassical energies (Eq. (\protect\ref{4.10.1}))
are drawn  by dashed lines for the lowest three  pseudorotational  multiplets
($n_\gamma$=0, $L$=0, 2, 4). The two-electron $S=1$ spectrum is
the same as for $n=1$, $S=\half$.}
\label{ExactS2}
\end{figure}

\begin{figure}
\caption{Exact spectrum for three electrons ($S=\half$).
The semiclassical energies, Eq.(\protect\ref{4.23}), are drawn  by dashed lines
for the lowest three  pseudorotational multiplets, $n_\gamma$=0,
$(L,k)$=(1,0), (3,2) (3,0). }
\label{ExactS3}
\end{figure}

\begin{figure}
\caption{Pair binding energy $U$ (thick solid line), compared to weak coupling
perturbation theory for $g<<1$ (dotted line) and semiclassical theory for
$g>>1$ (dashed lines). $U_n$ is found to be the same for $n=1,3,5$ electrons.
The Migdal--Eliashberg approximation $V$ (thin solid line) is also drawn for
comparison. $g^{{\rm C}_{60}} $ is the range of physical coupling strength for
C$_{60}$.}
\label{ExactU}
\end{figure}


\begin{thebibliography}{99}
\bibitem{com} One should not confuse the pseudo--rotations of Eq. (\ref{4.4}),
which describe propagating vibrational waves, with real rotations of the
molecule. In solid state those are assumed to be frozen out by crystal fields
at the temperatures of interest. Moreover, for a real rigid body, each moment
of inertia cannot exceed the sum of the other two \cite{llm}.
\end{thebibliography}
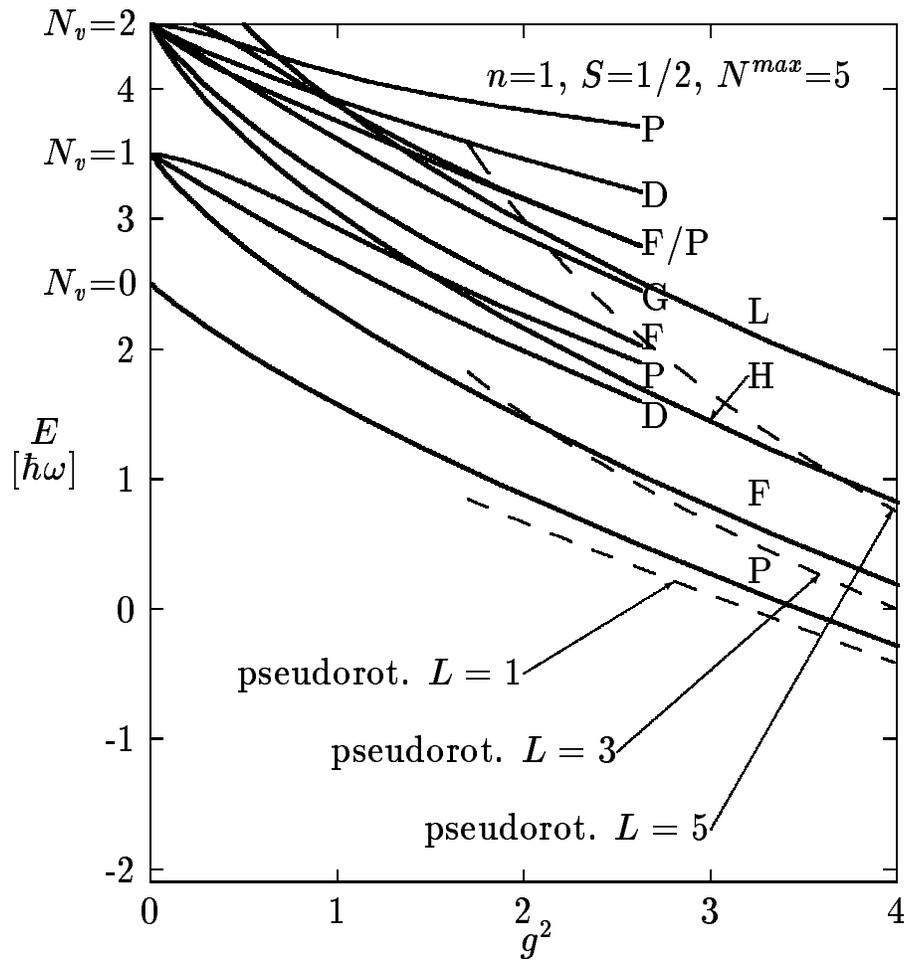
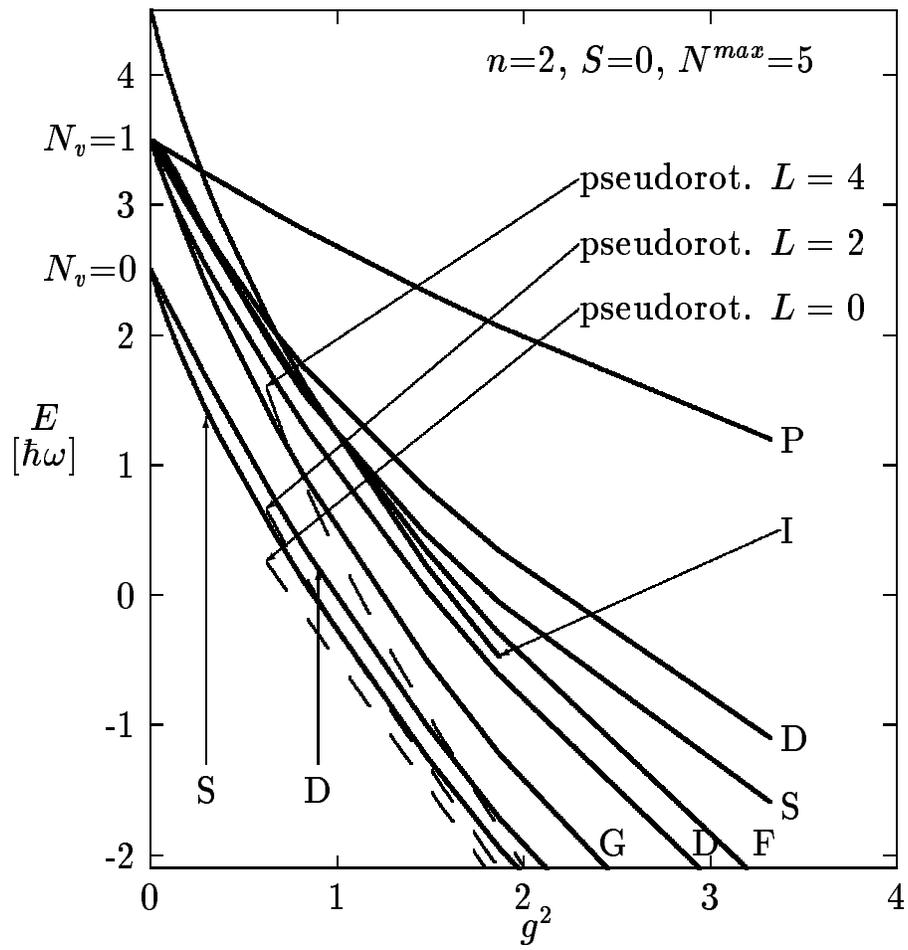
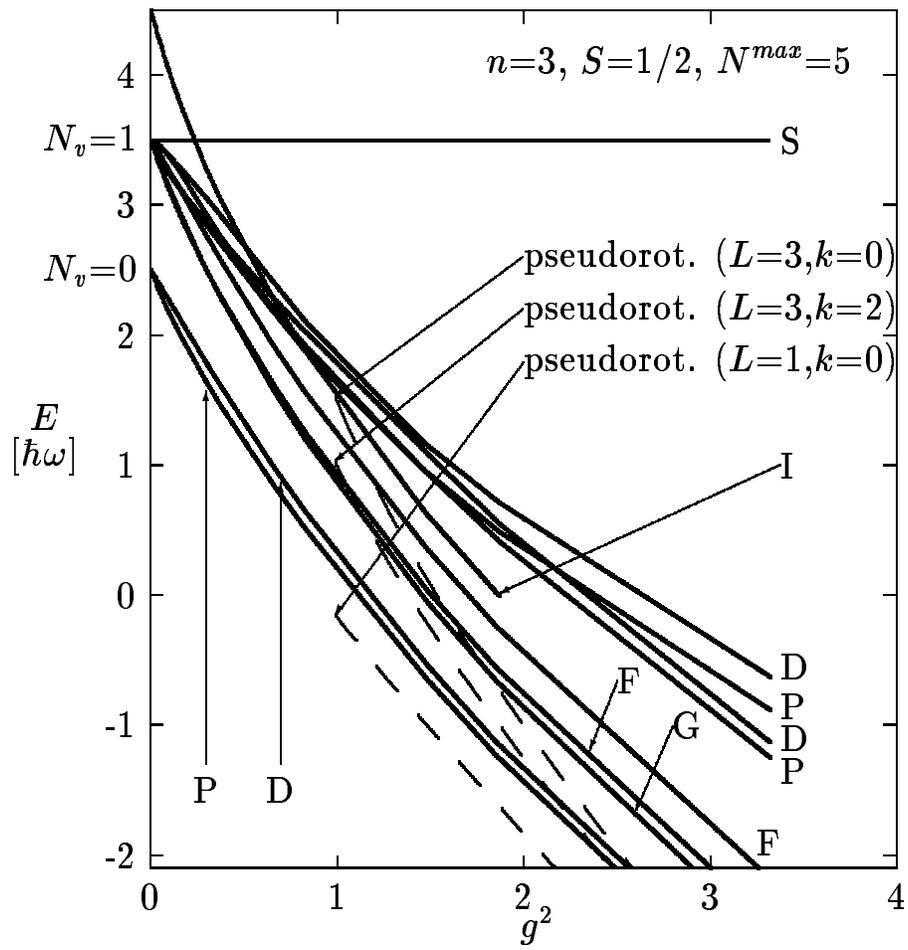
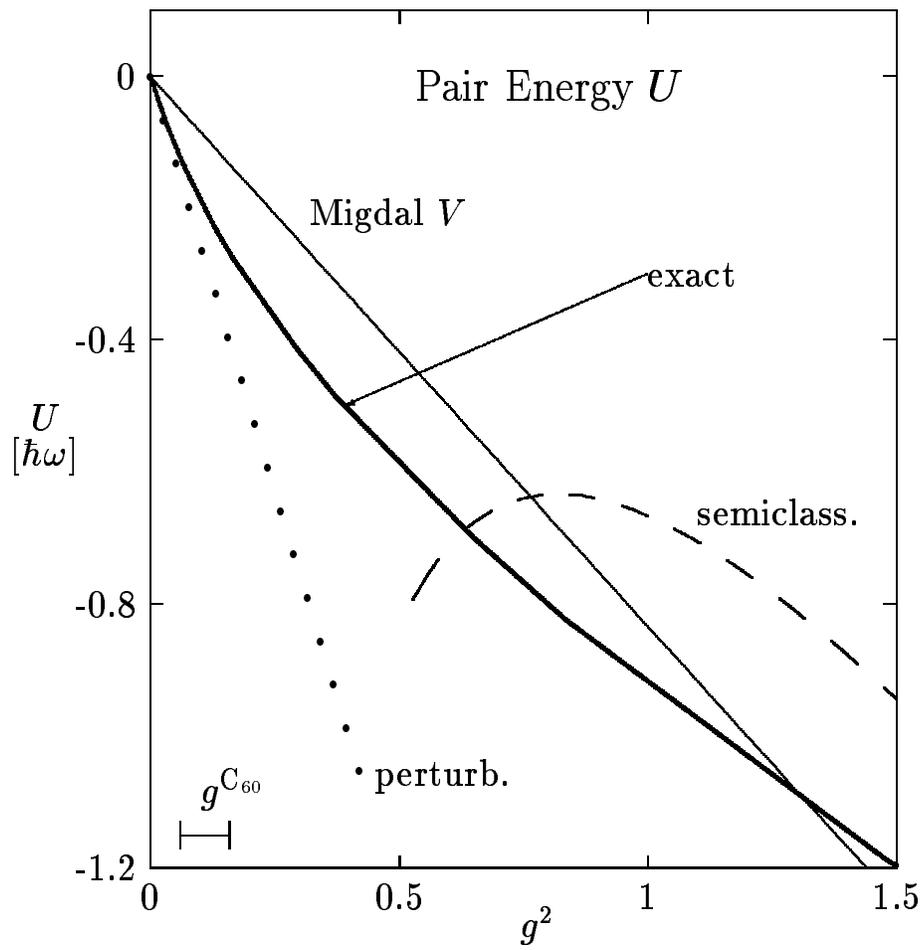
\end{document}